\documentclass[conference]{IEEEtran}
\IEEEoverridecommandlockouts
\usepackage{cite}
\usepackage{amsmath,amssymb,amsfonts}
\usepackage{graphicx}
\usepackage{textcomp}
\usepackage{xcolor}
\usepackage{algorithm}
\usepackage{algpseudocode}
\usepackage{amssymb}
\usepackage{balance}  
\usepackage{hyperref}
\usepackage{cleveref}
\usepackage{graphicx}
\usepackage{multicol}
\usepackage{pgfplots}
\usepackage{subfig}
\usepackage{tikz}
\usepackage{xspace}
\usetikzlibrary{arrows}
\usetikzlibrary{shapes.geometric,positioning}
\usetikzlibrary{patterns}
\usetikzlibrary{snakes}
\def\BibTeX{{\rm B\kern-.05em{\sc i\kern-.025em b}\kern-.08em
    T\kern-.1667em\lower.7ex\hbox{E}\kern-.125emX}}

\newcommand{\sysname}{KloakDB\xspace}
\newcommand{\mpc}{SMC\xspace}

\newcommand{\owner}{data owner\xspace}
\newcommand{\federation}{private data federation\xspace}
\newcommand{\hb}{query coordinator\xspace}
\newcommand{\shortsection}[1]{\noindent\textbf{#1}}
\newcommand{\kpv}{$k$-anonymous processing view\xspace}
\newcommand{\kqp}{$k$-anonymous query processing\xspace}

\newcommand{\ec}{equivalence class\xspace}
\newcommand{\qed}{\square}

\newtheorem{definition}{Definition}

\pgfplotsset{
/pgfplots/area legend/.style={
/pgfplots/legend image code/.code={
\fill[##1] (0cm,0.6em) rectangle (1*\pgfplotbarwidth,-0.3em);
}, }, }

\begin{document}

\title{KloakDB\\
Distributed, Scalable Private Data Analytics with $K$-anonymous Query processing
}

\author{\IEEEauthorblockN{Madhav Suresh}
\IEEEauthorblockA{
	\textit{Northwestern University}\\
	madhav@u.northwestern.edu}
\and
\IEEEauthorblockN{Zuohao She}
\IEEEauthorblockA{
	\textit{Northwestern University}\\
	zuohaoshe2013@u.northwestern.edu}
\and
\IEEEauthorblockN{Adel Lahlau}
\IEEEauthorblockA{
	\textit{Northwestern University}\\
	alahlou@u.northwestern.edu}
\and
\IEEEauthorblockN{William Wallace}
\IEEEauthorblockA{
	\textit{Northwestern University}\\
	williamwallace2018@u.northwestern.edu}
\and
\IEEEauthorblockN{Jennie Rogers}
\IEEEauthorblockA{
	\textit{Northwestern University}\\
	jennie@northwestern.edu}
	}

\maketitle

\begin{abstract}
	
	We study the problem of protecting side channel leakage
observed through query execution in a private data federa-
tions. Private data federations enable multiple data owners
to combine their data and query the union of their secret
data. Existing work focuses on providing data confidentiality
via secure computation and protecting side channel leakage
through variants of oblivious computation. This work uses
existing secure computation platforms and provides a novel
relaxtion of oblivious computation. While oblivious query
processing provides strong privacy guarantees, the signifi-
cant peformance overhead makes it infeasible for large scale
workloads.
We propose a novel semi-oblivious query processing model
for private data federations. We provide users with fine
grained privacy and performance tradeoffs for query execu-
tion. Our model, $K$-anonymous query processing, is based on
the widely deployed privacy model $k$-anonymity. $K$-anonymous
query processing ensures that the access patterns and sta-
tistical information revealed through query execution are
indistinguisable from those of $k$ or more records. The model
upholds strong privacy guarantees in the presence of multi-
ple query workloads. Our prototype \sysname is a decentral-
ized private data federation, showing adjustable speedups of
15X-1060X over oblivious query processing.

\end{abstract}

\section{Introduction}
\label{sec:intro}

People and organizations are collecting data at an unprecedented rate and many independent parties collect data on the same topic or area of research.  For example, consider two hospitals  serving a single patient but each keeps their own records on this individual.   Data sharing is crucial for enabling people to realize comprehensive insights from these fractured datasets.
On the other hand, {\owner}s are often hesitant to release their data to others owing to privacy concerns or regulatory requirements.  Some release anonymized versions of their datasets, but this compromises the semantics of the data in unspecified ways and this makes it difficult to join anonymized data with the data of others.  In contrast, a {\em private data federation} makes it possible for {\owner}s to pool their private data so that clients query and receive precise insights without needing anyone to disclose their sensitive records to others in the federation.  We consider this challenge in the context of a data federation, wherein multiple autonomous database systems are united to appear as one for querying.   A \federation performs privacy-preserving data analytics either using cryptographic protocols~\cite{bater2017smcql, bogdanov2008sharemind, conclave} or hardware enclaves~\cite{costan2016intel,kaplan2016amd} to combine the sensitive data of multiple parties.


In practice,  {\owner}s usually pool their private data with the assistance of a trusted  query coordinator, who collects the sensitive tuples of multiple parties, computes on them in plaintext, and sends the results to the client.   Hence, the {\owner}s never view one another's data and the client only accesses the output of the query, not any input tuples or intermediate results.   This pattern of relying on a  trusted \hb arises in many settings including financial regulators auditing banks~\cite{fcaTR, fincenCTR}, electronic health record research networks~\cite{fleurence2014launching, kho2014capricorn}, and prescription drug monitoring programs~\cite{finklea2014prescription}. However, relying on a trusted query coordinator brings downsides. First, finding a trusted third party can be challenging, and at times inappropriate given the domain. Second, this burdens the single trusted party with providing the computational resources required to compute queries on the joint data set. This setting does not enable {\owner}s to contribute their existing resources to query processing. Data owners desire a decentralized and distributed private data federation 
which enables {\owner}s to retain exclusive ownership over their raw records, while removing the trusted third party. 

Hardware enclaves as well as secure-multiparty computation alone solve the problem of decentralized 
confidential query processing for private data federations. They however do not protect against side channel leakage, specifically in the form of
 network traces, I/O patterns, memory access patterns, and statistical information leaked from operator trees. 


Existing work solves the problem of side channel leakage with oblivious 
query processing. Here the parties computing the query 
learn nothing more about the inputs of others than they 
would if if all of the members of the federation
uploaded their data to a trusted third party that ran the query and returned its results to the client. 
However the substantial overhead makes it impractical to 
use on large data sets or complex queries. 
For example a oblivious equi-join with fully padded outputs on 
two relations of length $n$ will unconditionally output $n^2$ output tuples. 
We note that while oblivious join algorithms exists which don't have $O(n^2)$ complexity , in order to mask the output cardinality and join correlations, the join 
must output the worst case cardinality unconditionally.
 Oblivious query processing's all or nothing approach is inflexible, not offering federation members 
any trade-offs for privacy and performance.  

There are several features that are desirable to have in a semi-oblivious query processing mechanism.  First, it should reduce the  computational complexity of a privacy-preserving query in comparison to full-oblivious evaluation.  Second, it should work efficiently in a distributed query processing setting.  Third, it should uphold its privacy guarantees over a wide range of SQL queries and repeated querying over a given dataset.   We introduce a semi-oblivious query processing model, \textit{$k$-anonymous query processing}, to address these goals.

$K$-anonymous query processing builds on the principles of $k$-anonymous data releases.   A data release is $k$-anonymous if each of its tuples is indistinguishable from those of at least $k-1$ others.  By generalizing this to query processing, we partition the input data into batches of tuples that we process obliviously, thereby reducing the overhead of multi-tuple operators like joins and aggregates.  In addition, tuning the $k$ parameter for a workload offers users a simple and intuitive ``knob'' with which to make trade-offs between privacy and performance. 
We also focus on $k$-anonymity because it is the de facto standard for data release and sharing in electronic health records~\cite{el2011methods,office2002standards}, educational research~\cite{daries2014privacy, seastrom2017best}, and more~\cite{garfinkel2015identification, fcasm2005sdc}.  Many of these settings use this technique to achieve regulatory compliance.   Hence, this model of semi-oblivious computation mirrors the needs of many real-world {\federation}s.



{\em \sysname} is a decentralized relational \federation that offers \kqp.  It evaluates its operators with secure computation to protect the confidentiality of data in flight.  We also use this technology to make a query's transcript semi-oblivious by evaluating a query's data obliviously one \ec at a time.    This system is agnostic to its cryptographic back-end and supports query evaluation in software -- with secure multiparty computation (SMC) -- or hardware with a trusted execution environment such as Intel SGX.    

\sysname implements \kqp by constructing a {\em \kpv} over the  tuples of its member databases.  This builds on the concept of $k$-anonymous data releases.  This view is $k$-anonymous with regard to the attributes that inform the control flow of the query.  This view breaks the data into equivalence classes such that each one contains $k$ or more records with respect to the control flow. It provides robust workload protection policies, preventing unauthorized information leakage, and provides a knob to trade off between privacy and performance. Although we focus on data federations, all of this system's techniques are readily applicable to a single data-owner setting, e.g., when an organization outsources their database's storage and query evaluation to an untrusted cloud service provider.

The main contributions of \sysname are: 
\begin{itemize}
	\item Formalizing an integrating \kqp into a relational data federation.
	\item Designing and implementing a prototype decentralized private data federation \sysname, which uses \kqp as it's semi-oblivious query processing model, while 
		utilizing secure-multiparty computation and hardware enclaves interchangeably as trusted backends. 
	\item Evaluating KloakDB on both synthetic and real world workloads. 
\end{itemize}

The paper is organized as follows: Section ~\ref{background} provides background, Section ~\ref{sec:overview} provides 
an overview and guarantees of our system, Section ~\ref{sec:kqp} formalizes \kqp, Section ~\ref{prototype-kloakdb} details our prototypte, \sysname.  Section ~\ref{evaluation} presents our experimental evaluation.  We continue with a survey of related work in Section~\ref{related} and conclude.

\section{Background}
\label{background}


\subsection{Side Channel Leakage}
Side channel leakage is the information that is leaked through 
a program's execution behavior.  Memory access patterns, network patterns, CPU instruction traces contribute to the execution behavior of a program. With query processing, output cardinalities and execution time are are examples of execution behavior which leak side channel information.  
In the honest-but-curious setting, corrupted parties will 
observe these side channels in order to infer sensitive information on the data. 
While these adversaries are considered \textit{passive}, these side channels can lead to powerful attacks on private data.
Xu et al \cite{side_channel_jpeg} showed that access pattern leakage can extract complete text documents and outlines of JPEG images.
Through repeated querying and with access to the query distribution and the output cardinalities, an adversary can determine secret attributes of individual 
records in a database ~\cite{kellaris2016generic}. 
On a macro level, query processing can reveal secret distributional information about the input data ~\cite{arasu2014oblivious}. 

\shortsection{Existing Techniques and Mitigations}
Cryptographic techniques such as \mpc, ORAM, and oblivious transfer 
exist to prevent access pattern leakage. 
These techniques add prohitive overhead to query execution, 
and given that operators output length and runtime depend on the 
input, require full padding of the output to remain 
fully oblivious ~\cite{chanDiffObliv2019}.
Arasu et. al. ~\cite{arasu2014oblivious} provide algorithms 
for oblivious query processing in time $O(n\log n)$. 
Our goal is to provide methods to reduce the overhead of these algorithms.

We now provide a brief overview of private data federations, privacy-preserving query evaluation, and $k$-anonymity.

\subsection{Private Data Federations}

A \federation enables multiple, autonomous database systems to pool their sensitive data for analysis while keeping their input records private.  It starts with a common set of table definitions against which the client queries.  The tables may correspond to data on a single host or they may be horizontally partitioned among multiple data owners.  We focus on the latter scenario in this work, although the system's core architecture is amenable to both.  This shared schema, including its functional dependencies, is known by all parties.

This schema is annotated with a security policy.   Each attribute in a federation's shared schema has a security policy that determines the information {\owner}s are permitted to learn when processing a \sysname query.  A column may be {\em public}, or visible to other {\owner}s.  All other columns are $k$-anonymous and their contents may only be revealed to other data owners when their values are indistinguishable from those of at least $k-1$ other individuals. 

Before the federation accepts its first query, it does a two-stage setup.   First, the data owners and other federation stakeholders work together to create a security policy based on the best practices of their domain and any applicable regulations and they initialize the common data model with this.  Second, a coordinator works with the {\owner}s to perform private record linkage~\cite{he2017composing} over individuals or entities that have records spanning two or more  databases so that each one resolves to a single identifier.

Each data owner wishes to keep their sensitive tuples private, but they are willing to reveal the results of queries over the union of their data with that of other federation members.   The query client receives precise results from their queries -- that are not $k$-anonymous -- by default.  The data owners may optional add a security policy for the output of analysis on their combined data, such as restricting the columns that are visible to the client or noising their results using differential privacy.   

\subsection{Privacy-preserving Query Evaluation}
Privacy preserving query evaluation aims to provide confidentiality, while also protecting against side channel attacks. 
In order to preserve confidentiality, KloakDB 
utilizes hardware enclaves and secure multiparty computation as 
the trusted backends. 
Trusted backends enable users to share data with untrusted parties and run computation secretly. 
Secure enclaves are enabled by hardware features, whereas 
secure multi-party computation requires no additional hardware and is enabled by cryptographic primitives. 
Secure multi-party computation is slower than hardware enclaves, but offer stronger security guarantees.  
\subsubsection{Secure Multiparty Computation}
SMC enables multiple parties to compute functions on secret data without revealing the underlying inputs. 
Given a shared function $f(x,y) = z$, two parties can share encrypted data $x,y$ and get the result of the function $z$. 
KloakDB uses EMP Toolkit~\cite{emp-toolkit} as our SMC backend, for the speed that it offers, as well as ease of interface. EMP Toolkit 
implements a \textit{semi-honest} two party protocol. Briefly, in the \textit{semi-honest} we expect users to faithfully execute the SMC protocol, 
however they will attempt to learn any information they can through observing the transcript of protocol execution. 
Our current SMC implementation allows for only two parties, and does not achieve feature parity with our hardware enclave implementation.
Recent work has demonstrated efficient SMC protocols for more than two parties ~\cite{global_mpc}, we leave it to future work to extend 
our implementation to utilize more than two parties.

\subsubsection{Hardware Enclaves}
Trusted execution environments such as Intel SGX~\cite{costan2016intel} and AMD Memory Encryption~\cite{kaplan2016amd} are available on most new commodity systems.    They may only execute  trusted code provided by the coordinator. The code and data associated with an enclave is {\em sealed}; the system in which it is executing may not view or change its contents.  
This hardware uses remote attestation to prove to an authority, such as the coordinator, that the code and data it is running has not been tampered with and that the code executes on  trusted hardware alone.    Once an enclave has attested its code,  this opens up a secure communication channel for the  {\owner}s to send their sensitive data to it.   
Secure enclaves have a protected region of memory, the encrypted page cache (EPC), that is not accessible by the host operating system or hypervisor.     
Recent research efforts have shown multiple vulnerabilities in hardware enclave implementations. 
Intel SGX's memory protection has received substantial interest from the security community, with side channel attacks being discovered ~\cite{costan2016intel,lee2017inferring,van2018foreshadow,wang2017leaky, van2019breaking} and fixed~\cite{costan2016sanctum,shih2017t, sasy2017zerotrace} on a regular basis.  Addressing the shortfalls of present-day hardware enclave implementations is beyond the scope of this work.

%

\subsection{K-anonymity}
\label{sec:background-k-anon}
A dataset is  $k$-anonymous iff each tuple in it is indistinguishable from at least $k-1$ others.  A {\em quasi-identifier} is the minimal set of attributes in a database that can be joined with external information to re-identify individuals in the dataset
with  high probability~\cite{samarati1998protecting}.     We focus on $k$-anonymity to protect sensitive data while providing high-performance query processing despite its security shortcomings when used for data release~\cite{aggarwal2005k, machanavajjhala2006diversity, martin2007worst, narayanan2008robust, sweeney2000simple}.   We do this because of the technique's potential for direct impact in many important application domains.  In addition, we observe that we use it to protect the transcript of a query's execution, rather than the results of its queries. We recognize that differential privacy has 
 surpassed $k$-anonymity as the golden standard for privacy release. However, we believe 
that \sysname can be useful for those domains and users which continue to utilize $k$-anonymity.

There is an abundance of research on constructing a $k$-anonymous release from a private dataset~\cite{bayardo2005data, el2009globally, lefevre2005incognito,lefevre2006mondrian,samarati2001protecting}.  They are optimized for {\em utility} and they minimize the difference between the released data and its private, original values.  This preserves the semantics of the source data  to make analysis on it as accurate as possible.  In contrast, with \kqp our goal speed up query runtimes.  The engine produces exact query results by default.  Hence, we assign the records to small, indistinguishable {\ec}es  that minimizes the size of each bin to reduce our overhead  while upholding the system's privacy guarantees as tuples move up the query tree.  


\section{Overview}
\label{sec:overview}

We now lay out the preliminaries for \sysname. First, we define the system's trust model and security guarantees.  We then describe the architecture of this \federation and walk through the steps with which a query will run in this setting.    We then introduce a running example.

        

\subsection{Private Data Federation}
\shortsection{Privacy Assumptions}
We assume an honest-but-curious private data federation. This is considered to a standard trust model for the untrusted cloud setting ~\cite{mehrotra2019querybinning}. 
In the PDF setting, the data owners are the potential adversaries. We trust the {\owner}s to  run the SQL statements from the query plan to provide accurate inputs for  evaluation within the secure computation.  On the other hand, each {\owner} will attempt to learn as much as they can from observing the query's execution. {\owner}s will monitor memory, CPU, and network side channels, as well as timing and intermediate result sizes with local and distributed operators. 

If we are executing in trusted hardware, we need not trust the {\owner}s in \sysname to faithfully execute query operators over the sensitive data of others.   The enclaves ensure that queries over the data of others are completed with integrity and without unauthorized access to their values.   Thus even if the {\owner}'s operating system is compromised, or they attempt to act maliciously they will not succeed in compromising the security policy of a \federation.   For running with \mpc, we operate in the semi-honest model -- trusting the {\owner}s to faithfully execute these cryptographic protocols.   Likewise, the \sysname query planner is trusted to only admit queries that meet the security policy of the \federation and to construct query plans that will uphold those policies. A client's query is visible to all parties including the {\owner}s.

The \sysname private data federation consists of two parties:
\begin{enumerate}
	\item \textit{Client}: The client has access to the shared schema that the federation has agreed upon. The client can issue 
		any query to the federation. The client recieves an encrypted query result from the federation, and can observe the amount of time the query takes,  however cannot see intermediate results. 
	\item \textit{Data Owner}: Data owners agree upon a shared set of table definitions a priori. All query processing and computation 
		happens within the pooling of data owner's resources. Data owners recieve encrypted secret data from each other. They have 
		access only to their own unencrypted data. Through \mpc or trusted hardware they are able to compute over encrypted data 
		shared by other data owners, while combining their secret data. Data owners will snoop on all side channels. 
\end{enumerate}

\subsection{Problem Statement}
We consider a private data federation $\mathcal{P}$ consisting of $n$ data owners $\mathcal{D} = \{D_1, \ldots, D_n\}$ with a shared 
schema $\mathcal{T}$ consisting of relations $\mathcal{R} = \{R_1, \ldots, R_n\}$. The data owners wish to issue a query workload 
$\mathcal{Q} = \{Q_1, \ldots, Q_n\}$. We assume the data owners fit the honest-but-curious security model. Data owners have the following requirements: 
\begin{enumerate}
	\item To pool together their computational resources for query processing.
	\item To maintain confidentiality of their private data. When unioning their data for querying within the federation, they will only share
		encrypted data.
	\item To be protected from side channel leakage throughout the query workload. 
\end{enumerate}


\subsection{Security Guarantees}

We now introduce the anonymity guarantee \sysname offers  {\owner}s during query execution.  These queries have instruction traces that have the same distribution as they would if the query were executing over a view of the dataset that is $k$-anonymous with respect to attributes that alter its control flow.   {\em Control flow attributes} contain values that change the observable instruction traces of $k$-anonymous query operators running with secure computation, e.g., how it branches, loops, and their result sizes.   We  formally introduce them in Section~\ref{sec:kqp} and denote the list of attributes with $C$.    With this in mind, \sysname guarantees: 

\begin{definition}
	{\bf $K$-anonymous Query Processing}
	Given a set of queries $\mathcal{Q} = \{q_i\}$ that access relations 
	$\mathcal{R} = \{R_i\}$, defined by the schema $\mathcal{T}$.
	$Q$ has control flow attributes $C$, such that $C \subseteq \mathcal{T}$. When the system evaluates $Q$ on $\mathcal{R}$, there exists a function  
	$\mathcal{V}_C$ for creating a $k$-anonymous view of $\mathcal{R}$ with respect to $C$. A semi-oblivious 
	algorithm $\mathcal{A}$ for running the query workload, satifies the requirements of $k$-anonymous query processing iff its instruction 
	traces are computationally indistinguishable from those of a simulator running $\mathcal{Q}$ over $\mathcal{V}_C(\mathcal{R})$ for 
	a probabilistic polynomial time adversary. Therefore:
	$$Trace(Sim(\mathcal{V}_C(\mathcal{Q}))) \stackrel{{\rm c}}{\equiv} Trace(\mathcal{A}(\mathcal{Q}))$$
\label{def:kqp}
\end{definition}
\vspace{-5mm}

\begin{definition}
\end{definition}

Intuitively this definition states that any algorithm over a query workload which guarantees that the instruction traces of a single tuple are 
indistinguishable from at least $k-1$ other tuples satisfies $k$-anonymous query processeing. We introduce the notion of a $k$-anonymous view in Section ~\ref{def:k-anon-view}.


\begin{figure}
    \centering
    \includegraphics[width=0.5\textwidth]{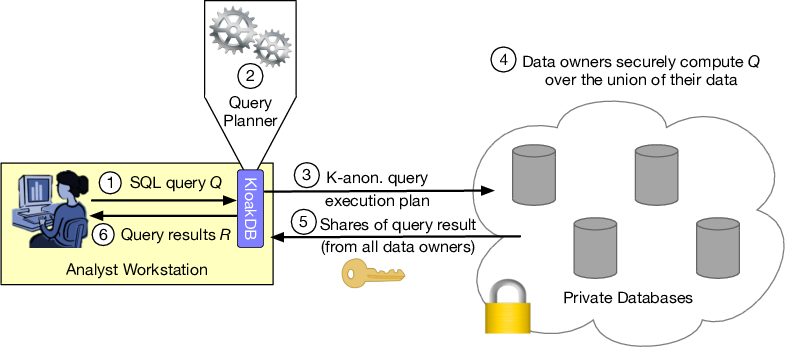}
    \caption{KloakDB query workflow.  Lock denotes secure computation and the key labels an encrypted communication channel.}
    \label{fig:workflow}
\end{figure}

\subsection{Query Workflow}

We examine the workflow of a \sysname query in Figure~\ref{fig:workflow}.  

\textbf{1.} It begins when a client submits a SQL statement, $Q$, to the system's thin client running on his or her machine.  The client may be a data owner in the federation or an external user, such as a researcher in the hospital example in Section~\ref{sec:intro}.    \sysname parses the statement into a directed acyclic graph (DAG) of database operators, its query tree.   

\textbf{2.}Next, the client-side planner analyzes the query tree and uses the schema's security policy identify the minimal subtree of its DAG that requires privacy-preserving computation.   The planner assigns each node to run in  plaintext or $k$-anonymous mode.   It also identifies the query plan's control flow attributes with which it will select a \kpv with which to run its secure operators.  More details on this step are in Section~\ref{sec:kqp}.

\textbf{3.} The query planner next translates the query into a $k$-anonymous query execution plan.   For the operators computing on public attributes, we run them in the clear.  Whenever possible, we run them locally and if we do distributed computation in the clear, we project out any private attributes first and re-join with the base tables after this.  The planner takes the plaintext subtrees and generates a SQL statement for each and the {\owner}s run them within using their DBMS.   \sysname then generates secure computation protocols for the operators that will run $k$-anonymously.  It works with templated code for each operator, parameterizing it with the fields they will access, predicates, and other expressions.  If the query is running in trusted hardware, the planner creates a query template runs it on pre-compiled secure enclave code.   If we are using SMC then it translates the operator into secure multi-party computation primitives.   At the time of this writing, \sysname supports a query using SMC or secure enclaves but not a hybrid of the two.   



\textbf{4.} Now that we have a secure query plan that upholds 
our privacy guarantees, the \sysname client sends it to the 
data owners. The query plan is public to all data owners. 
The data owners run the secure plan's input SQL statements
in parallel. They now switch to $k$-anonymous mode. We 
describe the details of $k$-anonymous query processing 
in Section ~\ref{prototype-kloakdb}.

\textbf{5.} After the query is securely computed amongst the data 
owners, the encrypted results are sent back to the \sysname client.
The \sysname client decrypts and assembles the query results - removing 
any dummy tuples - and returns the final output to the client who issued
the query.

\subsection{Running Example}
\begin{figure}
\centering
\includegraphics[width=\linewidth]{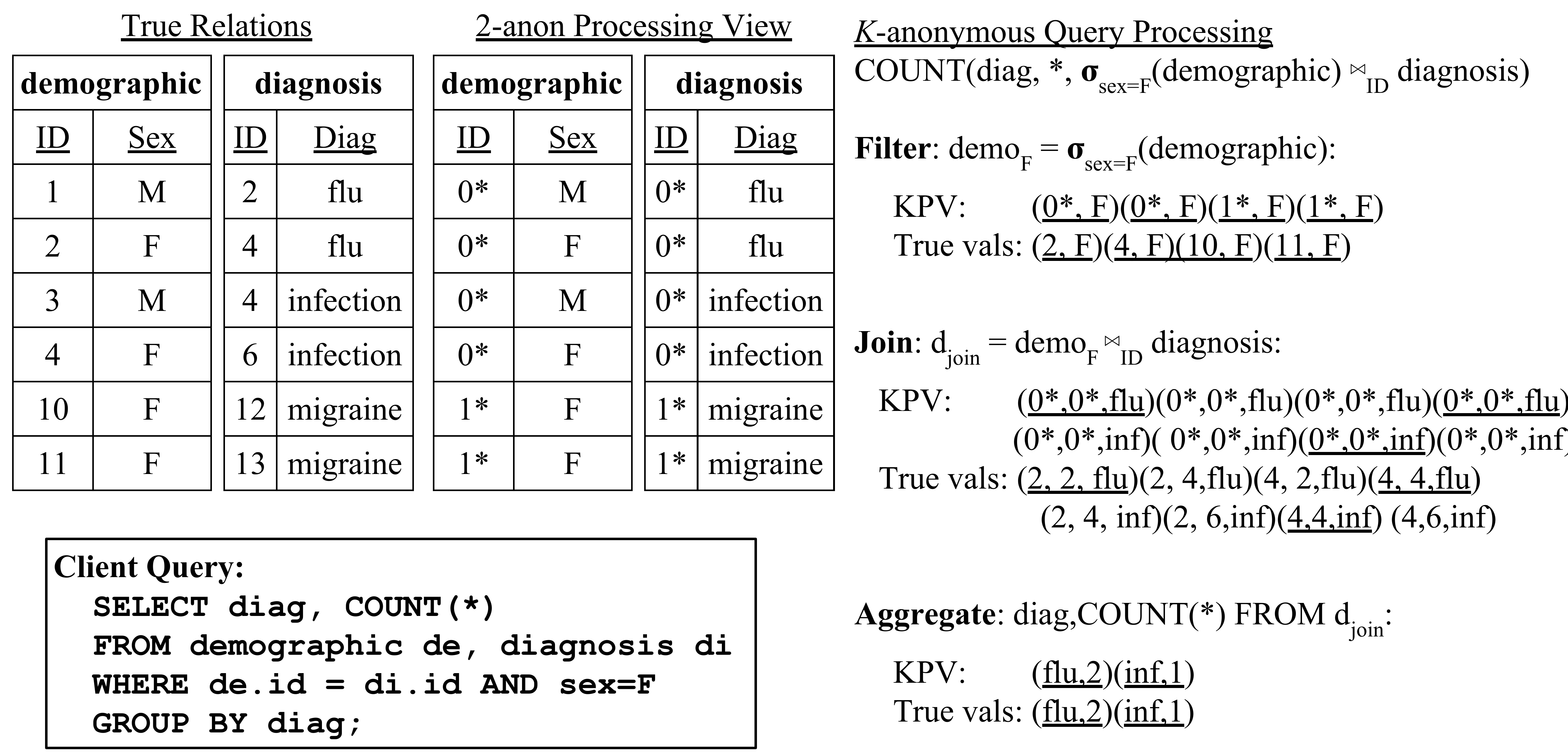}
\caption{$K$-anonymous query processing example.  True (non-dummy) tuples are underlined. }
\vspace{-5mm}
\label{fig:k-anon-query}
\end{figure}
\label{sec:running-example}

Consider the query  in Figure~\ref{fig:k-anon-query}. It counts the times a woman is diagnosed with a given ailment.  It first filters the demographic table for women, joins the selected tuples with the diagnosis table, and  counts the times each condition appears in the join result.    Sex and diagnosis have a $k$-anonymous security policy where $k=2$.  We demonstrate this in the single database setting such as that of outsourced operations in the cloud.  

 The query planner first creates a $2$-anonymous processing view.   To illustrate this, we use generalization\textendash omitting the least significant digit of the IDs\textendash instead of \sysname's freeform $k$-anonymization here.   The view must have at least two individuals (IDs) in each equivalence class.  All of the tuples in an \ec are indistinguishable from one another during query processing.  We divide the ID column by 10 in both relations to suppress the least significant digit.  Each relation has three {\ec}es, and for demographic they are: (0*, F) (0*, M), and (1*, F).   Each equivalence class has a bitmask with a bit or dummy tag for each tuple denoting if it is a placeholder to mask the role of individual tuples in the group.  When we run the query, the filter first examines each \ec  and either  1) outputs it in its entirety if it contains at least one match\textendash  obliviously marking a dummy tag on each tuple to denote if it met the selection criteria; or 2) produces an empty set.   The filter outputs two  of the three demographic {\ec}es.
 
 Next, we join the filtered demographics tuples with the diagnoses using the same all-or nothing logic to uphold tuple indistinguishability  over an \ec.  When the join compares two {\ec}es its output is either size of their cross-product of its inputs or an empty set.   This join outputs  two {\ec}es: (0*, flu) and (0*, infection) and three true tuple matches.   If we ran this obliviously  the join would output the cross-product or 36 tuples, instead of the 8 shown here. Clearly, \kqp has an opportunity to substantially boost the performance of \federation queries.

 After the join, the aggregate iterates on its results  one \ec at a time to count up the diagnoses for each ailment.  For a given group-by bin, an aggregate outputs either: 1) a single tuple if $>=k$ individuals contributed to it; or  2) a dummy-padded set of tuples equal in length to the source \ec.  An observer can learn about no fewer than $k$ individuals at a time by observing these outcomes because they either learn that all of the tuples in the class had the same group-by value  or that we processed the \ec obliviously.   At first glance, it may appear that the group-by of (0*, infection) would be processed obliviously.  As we will see in the coming sections, the anonymity of an \ec is transitive as it passes through a $k$-anonymous operator. Because the join compares all tuples in an \ec to its potential matches in the joining relation, its output is fully padded.  Hence, the join did not reveal its selectivity over this \ec and the groups with which it was paired.  Then the count operator visits all four tuples in the infection group and emits a single tuple with the true count.  

\section{$K$-anonymous Query Processing}
\label{sec:kqp}

We now reveal how \sysname upholds the anonymity guarantees in Definition~\ref{def:kqp}.  We do this using the notation in Figure~\ref{tab:kqp}.   Consider a query $Q$ that evaluates over the tables in the \federation's shared table definitions, $\mathcal{R} = (R_1, \ldots R_n)$.   This schema is a lossless, dependency-preserving join decomposition. Intuitively we require the schema to be a lossless join decomposition since it allows us to view the union of the relations as a single "super-relation". 
We found this idea useful for reasoning about the properties of $k$-anonymity within multi-relational schemas. We note that not all schemas 
are lossless, dependency-preserving join decompositions, thus this requirement restricts the setting for $k$-anonymous query processing. We query a database instance, $\mathcal{D}$.   To capture the information revealed as tuples flow up the query tree, we model our query processing in terms of a schema-level $k$-anonymous view  comprised of natural joins as  $R_* = R_1 \bowtie \ldots  \bowtie R_n$.  
Recall that in a conventional $k$-anonymous data release each tuple in a database instance $\mathcal{D}$ is indistinguishable from at least $k-1$ other tuples with respect to its quasi-identifiers.

We first determine the control flow attributes that must be handled with $k$-anonymity for $Q$.  First, we protect the quasi-identifiers  defined by the federation's access control policy by anonymizing the ones that alter the control flow or intermediate result sizes of our query operators.  Second, we take into account the attributes that modify the control flow of any operator that follows an anonymized one.    For example, the query in Figure~\ref{fig:k-anon-query} joins on a public attribute -- patient ID -- and it does so after a filter on a quasi-identifier.  In order to not reveal the patient IDs that are women, we anonymize with respect to both sex and patient ID.


 {\em Control flow attributes} are the union of the schema's quasi-identifiers and the attributes that alter the control flow of $Q$ after computing on a quasi-identifier. Recall that we denote them as $C$.  $V_C$ a $k$-anonymous view of $R_*$ with respect to $C$.   The attributes that alter the control flow of each  database operator type are well-known.  For joins, their transcript is based on the values in their join keys, filters base their control flow using their predicates, aggregates with their group-by clauses, sorts on their sort keys, and set operations on all columns on which they are evaluated.

\begin{definition}  {\bf Multi-relation $k$-anonymity} Consider a $k$-anonymous view, $V_C$ over relations $\mathcal{R}$ that is anonymized wrt attributes $C$.  We say that this view of a database instance $\mathcal{D}$ is $k$-anonymous iff for every valid value  $t_i \in C$, $\sigma_{C=t_i}(V_C(R_*))$  produces either  $\geq k$ tuples -- that may include duplicates --  or an empty set.
\end{definition}





In order for us to maintain a $k$-anonymous view of $\mathcal{D}$ among the data owners, an execution  transcript of $Q$ running over $\mathcal{D}$ may reveal no more information than we could glean from observing an execution of $Q$ over $V_C(R_*)$.   Individual $k$-anonymous operators in \sysname do this by obliviously evaluating over each \ec discretely.  This upholds the $k$-anonymous view among the {\owner}s owing to the following property.

\shortsection{Subset Property}
\label{subsetprop}
\cite{lefevre2005incognito}:  If $R$ is $k$-anonymous with respect to $C$, then $R$ is $k$-anonymous with respect to any set of attributes $P$ such that $P \subseteq C$.\\
\noindent {\bf Proof:} Consider the frequency set of $R$ with respect to $C$. If we remove any attribute $C_i$ from $C$, then each of its equivalence classes will remain the same, or it will coalesce with another one. Thus each frequency set will be greater than or equal to its previous size. $\qed$

In other words, an operator with a control flow that is a subset of $C$ will merge {\ec}es from the initial view and thus still be $k$-anonymous in its query processing.

In addition, we need to ensure that as we sequentially run operators in the query tree that composing them will uphold our security guarantees:

\shortsection{Transitivity Property}:  Given a relation $R$ that is $k$-anon\-ymous with respect to $C$, the execution and output cardinalities of any transformations predicated on $C$ or $P \subseteq C$ are themselves $k$-anonymous.

\shortsection{Proof}: In $V_C(\mathcal{D})$ each tuple is indistinguishable from at least $k-1$ others.  Thus the transcript of transformations on a $k$-anonymous relation cannot reveal information that is not present in the source view. $\qed$

Owing to the transitivity property, we reason about the view with which we input each relation, $R_i$ into $Q$ according to its subset of the control flow attributes.  Since every $k$-anonymous operator that computes on $R_i$ will leak information about its control flow attributes or a subset thereof, it will uphold the federation's anonymized view of the data.  Hence for all $R_i \in Q$, we create a {\em $k$-anonymous processing view}.

\begin{definition}
\label{def:k-anon-view}
{\bf $K$-anonymous Processing View} A relation $R_i$ has control flow attributes $c_i = R_i \cap C$.  The relation is anonymized as $R_i' = V_{c_i}(R_i)$.  $R_i'$ is suitable for \kqp  iff for all possible projections, ${\LARGE \pi}_{P}(R')$, where $P \subseteq c_i$ produce $\geq k$ tuples.  Its output admits duplicate rows. 
\end{definition}

When we compute over a \kpv, we run a query $Q$ over a subset of the relations, $Q(v_{c_1}(R_1) \bowtie \ldots \bowtie v_{c_n}(R_i))$.  Since we are eagerly anonymizing the control flow attributes, our execution traces will protect at a level greater than or equal to that of running $Q(V_C(R_*))$.

Before we describe how \sysname's operators provide the invariants above, we extend {\kpv}s to the  federated setting.  When considering anonymized views in Definition~\ref{def:kqp}, the data is not combined with tuples from other hosts. In a \federation a curious data owner may analyze a queries instruction traces to infer information about the secret inputs of their peers by ``subtracting out'' their contribution to $Q$. 

We now generalize  Definition~\ref{def:k-anon-view}  to {\federation}s to support the partial view of each \owner that evaluates one or more equivalence classes in query $Q$.    This unified view ensures that each host participating in query processing learns about no fewer than $k$ tuples at a time when they run operators over an anonymized view.  

\begin{definition}
{\bf Federated $K$-anonymous Processing View} $\mathcal{D} = V_C(R_*)$ is horizontally partitioned over $n$ hosts, $\mathcal{D} = \{\mathcal{D}_{1}, \ldots, \mathcal{D}_{n}\}$.   To ensure that no \owner learns about fewer than $k$ tuples at a time, for all of the data owners, $i = 1\ldots n$, for all $P \subseteq C$,  ${\LARGE \pi}_P(V_C(\mathcal{D} - \mathcal{D_i}))$,  produces either $\geq k$ tuples or an empty set.  Its output rows may include duplicates.
\label{def:federated-k-anon-view}
\end{definition}

If a \kpv satisfies Definition~\ref{def:federated-k-anon-view}, then it will also uphold the guarantees of \kqp regardless of how much data was contributed by each host for a given operator.  This is because even if the host does a what-if analysis of removing his or her tuples from the equivalence class, it will not expose data about fewer than $k$ tuples.

\noindent{\bf Discussion} 
In the extended version of this paper we discuss the operator implementations. 
Briefly, all operators utilize modern oblivious techniques for execution while also
outputting fully padded cardinalities. With joins this is the full cross product, 
for filters, the input relation marked with secret dummy tags, and aggregates 
an unconditional tuple for each distinct group by attribute in the input relation. 
The key, and simple idea of this system is the reduce the unit with which oblivious computation is performed.  
$K$-anonymous query processing offers a rigorous approach to perform oblivious computation over smaller equivalence classes and it adheres to a well understood and widely deployed privacy model. Our approach does not provide any guarantees for data release, the domain in which $k$-anonymity is 
usually deployed.  Without operators outputting fully padded results the overhead for an oblivious operator algorithms is $O(n \log n)$ with respect to non-secure 
algorithms ~\cite{arasu2014oblivious}. Note that in the case of $k$-anonymous query processing, the input size to each oblivious operator algorithm 
is $k$ instead of $n$, since we are grouping tuples together with size $k$. Thus performance of any polynomial state of the art oblivious algorithm
will improve, since we decrease the input size into the oblivious algorithms. (We note that there will still be a multiplicative factor of $n/k$).


\section{KloakDB}
\label{prototype-kloakdb}
We built a prototype implementation of a data federation that uses $K$-anonymous query 
processing as it's semi-oblivious query processing model. 
To provide confidentiality for users data, the prototype 
allows for both SMC and secure hardware backends. 
Critically, our prototype requires no trusted third party: 
 KloakDB's query processing environment is entirely decentralized. 

 With an agreed upon workload, data owners can collectively agree upon 
 a set of control flow attributes $C_{init}$ and $k_{init}$. $(C_{init}, k)$ is 
 shared among the data owners, and the trusted comptuation platform is bootstrapped
 across the federation. When the 
When the trusted computation backend is SMC, this involves
bootstrapping the SMC library, when the backend is secure hardware, 
this involves sharing keys, booting up enclaves, and remote attestation. 
If $C_{init}$ is non-empty, KloakDB runs view anonymization on the 
input relations with $(C_{init}, k_{init})$. $(C_{init}, k_{init})$ is saved on all data owners 
as $(C_{system}, k_{system})$, which is used for workload protection. 
This concludes the system setup, 
now the federation is ready to receive queries from any data owner in the federation. 

\subsection{K-Anonymous View Generation}\label{view-generation}
In this section we discuss k-anonymous anonymized view generation.  
A random \textit{coordinator} is chosen amongst the data owners to run the
view generation process. 

\textbf{Step 1 Statistics}
First the coordinator requests encrypted histograms for the relations 
grouped by the control flow attributes, over all data owners into their trusted backend. With hardware enclaves, the histograms are sent directly into the enclave, with 
SMC, the histograms are shared secrets between the two parties.  
Data owners sort the histograms by most frequently occurring values. Our security assumptions ensure that data owners will honestly send their statistics. 

\textbf{Step 2 View Generation}

\vspace{1mm}
\hrule
\vspace{.5mm}
\textbf{Anonymized View Generation Algorithm}
\vspace{.5mm}
\hrule
\noindent
\vspace{.5mm}

\textit{Input:} $K$-anonymous parameter $k$, Control Flow attributes $\{C_f\}$ to anonymize over, set of relations $\{R_i\} \subset \mathcal{R}$, statistics on relations $\{S_i\}$. 

\vspace{1mm}

\textit{Output:} Mapping $M : \{C_f\} \rightarrow EquivalenceClass_{ID}$

\vspace{1mm}
\hrule
With the histograms in hand, the coordinator runs an algorithm 
such that the views generated fit the definitions from Definition ~\ref{def:federated-k-anon-view}.
Any such algorithm which outputs views with those constraints will allow for 
$k$-anonymous query processing. 

\textit{View Anonymization Algorithms:}
The flexibility of view generation is a similar feature that exists in 
$k$-anonymous data releases. $K$-anonymity allows different algorithms 
to be used to create $k$-anonymity in a dataset, each with it's own objectives ~\cite{stat_trade_off_gen, doka2015k,nergiz2009multirelational, bayardo2005data, el2009globally, lefevre2005incognito,lefevre2006mondrian,samarati2001protecting}.
Possible objectives for users to optimize can be dependent on workload, 
numbers of nodes in the federation, and type of secure platform backend being used.  
Our prototype implements an algorithm similar to the greedy variant found in Doka et al. ~\cite{doka2015k}, extended to the multi-relational setting as in Nergiz et al. ~\cite{nergiz2009multirelational}. The algorithm seeks to minimize the maximum 
size of any equivalence class.


\textbf{Step 3 Mapping/Partitioning}
Partitioning has two different implementations depending on the trusted backend. This is the last stage of anonymization.

\textit{SMC:} We support two parties, and data is a shared secret resident on both machines, and no partitioning
or redistribution occurs. The generated map between control flow attributes and equivalence classes is shared across the two hosts, and each tuple privately shared across the two hosts, then mapped to the appropriate \ec in SMC. This ensures that the mapping remains unknown to the two hosts.
With respect to partitioning, the SMC model presents the union of the data as a single host, so operator execution behaves as it would in the single host setting. 

\textit{Hardware Enclave:}
After the algorithm has completed, the coordinator has a map between 
hashes of control flow groups and \textit{equivalence class IDs}. The coordinator 
sends the map, as well as a randomly generated hash function to each data owner's hardware enclave, where the data owners use the map to create local anonymized views of the data. Using the hash function from the coordinator, 
each data owner $D_i$ hashes the equivalence class ID to determine which 
data owner, $D_r$ should receive the local anonymized view. Then the data owner $D_i$ securely sends the equivalence class to the hardware enclave of the receiving data owner $D_r$.  $D_r$ then combines the equivalence classes
from all sending hosts to form a local partition of the anonymized view.

\subsection{Query Workflow}
Queries are requested into KloakDB by clients as a pair, $(Q, k_Q)$, with $Q$ a query, and $k_Q$ the associated 
$k$-anonymous security parameter. Note, data owners explicitly are the only clients allow to query the federation. 
First the query $Q$ is decomposed to determine it's control flow attributes $C_{Q}$. 
$C_{Q}$ is compared against $C_{system}$, if $C_{Q} \subseteq C_{system}$, and $k_Q \leq k_{system}$, 
the query proceeds with $(C_{system}, k_{system})$. Through the subset property in Section ~\ref{subsetprop}, we will still maintain $k$-anonymous query processing.

If $k_{Q} \geq k_{system}$, and $C_{Q} \subset C_{system}$, the equivalence classes present in 
the anonymized views must be combined to 
satisfy the security requirements for the query. We run a simple greedy algorithm 
to combine equivalence classes until they satisify $k_{Q}$, then let $k_{system} = k_{Q}$. 
If $C_{system}$ and $C_{Q}$ are disjoint, then the system takes the union, $C_{system} \cup C_{Q}$, and runs view augmentation. If $C_{Q} \not \subset C_{system}$, the system upgrades the query 
to full oblivious computation. The key is for a set of queries that run over a set 
control flow attributes, a single $k$-anonymous view can be used multiple times without 
any privacy loss.

\begin{figure}[t]
\centering
\resizebox{0.4\textwidth}{!}{
\begin{tikzpicture}[snake=zigzag, line before snake = 5mm, line after snake = 5mm]
\draw (0cm,5pt) -- (0 cm,-5pt);
\draw (5cm,5pt) -- (5cm,-5pt);

\draw (7cm,5pt) -- (7cm,-5pt);

\draw (0,0) node[below=5pt] {$ k=1 $} ;
\draw (0,-0.35) node[below=5pt] {Encrypted} ;
\draw (0,-0.7) node[below=5pt] {(fastest)} ;

\draw (5,0) node[below=5pt] {$ k=n/2 $} ;

\draw (7,0) node[below=5pt] {$ k=n $} ;
\draw (7,-0.35) node[below=5pt] {Oblivious} ;
\draw (7,-0.7) node[below=5pt] {(slowest)} ;

\draw (0,0) -- (5,0);
\draw[snake](5,0) -- (7,0);
\draw [thick,decorate,decoration={brace,amplitude=6pt,raise=0pt}] (0,0.3) -- (5,0.3);


\node[align=center] at (2.5,0.75) {$K$-anonymous query processing};

\end{tikzpicture}
} 
\caption{The privacy-performance decision space offered by \kqp.}
\label{fig:continuum}
\vspace{-5mm}
\end{figure}
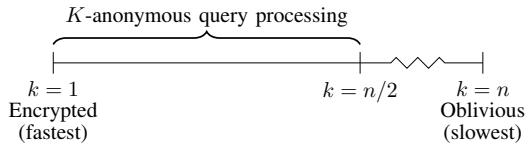

\subsection{Privacy-Performance Trade Off}
\label{sec:knobs}

 $K$-anonymous query processing enables federation members to achieve a profitable trade off between performance and privacy.  We visualize this decision space in Figure~\ref{fig:continuum}. Consider a medical researcher querying their electronic health records that are stored using encryption in the cloud.  She wishes to set her $k$ to a higher value when she is querying highly sensitive data.   For example, many states require records pertaining to the treatment of HIV and other sexually transmitted infections have greater $k$-values than more common diagnoses~\cite{gostin1996public}.  When accessing these records, she would use oblivious querying.  For more common ailments, she is willing to forgo stronger privacy guarantees in exchange for faster query runtimes.

This decision space, a range of $k$ values for anonymization, arises in many settings.   In clinical data research, guidelines for $k$-anonymization vary.  A $k$ from  5-11 is recommended for most health contexts~\cite{klein2002healthy,nshs2012disclosure,pcornet2016governance}, although some data providers suggest $k=3$~\cite{lo2015sharing} and other, more sensitive studies call for $k=30$~\cite{cdc2017yrbss}.  For educational data, the US's FERPA has various $k$-anonymization guidelines for a variety of data release scenarios in~\cite{daries2014privacy,seastrom2017best}.   Energy data is also has a range of $k$ values for its release from $k=5$~\cite{dcEnergy} to $k=15$~\cite{illinois1515}.  

By tapping into the expertise of the data federation, we will realize substantial performance gains by adjusting $k$ to the sensitivity workload at hand.   In practice, a \federation may have  heterogeneous security policies on client queries to address these domain-specific nuances. 

\section{Experimental Results}
\label{evaluation}
\subsection{Implementation}
KloakDB is implemented as an in memory distributed query execution engine.
We implement KloakDB in 4000 SLOC of C++. 
Our prototype supports uses Intel SGX as the hardware enclave backend 
and EMP-Toolkit ~\cite{emp-toolkit} as the secure multiparty computation library. 
 With EMP-Toolkit our implementation can support two data owners, however with SGX, it can support an arbitrary number of data owners.
Our prototype takes as input a SQL query and runs it through a standard SQL parser. After parsing the query is automatically decomposed to find the control flow attributes, after it anonymizes which anonymization occurs. The implementation uses off the shelf RPC libraries with SSL, and encrypts all data that occurs outside of the trusted backends. 
\subsection{Experimental Setup:}
We run experiments on two testbeds. Our hardware Intel SGX benchmarks 
are run on 4 Ubuntu 16.04 servers running Intel Core i7 7700k processors, 
with 32 GB RAM, and a 1 TB 7200 RPM HDD. The SMC experiments are run 
on two machines from the same test bed. Our benchmarks utilize KloakDB in four modes of query processing: \textit{plain}, \textit{encrypted}, \textit{k-anonymous}, \textit{oblivious}. Encrypted query processing mode does not run the queries obliviously, and does not pad the intermediate result sizes. Oblivious query processing mode runs the queries obliviously and fully pads the output sizes. \textit{Plain} mode runs using PostgreSQL's Foreign Data Wrapper (FDW) ~\cite{postgresFDW} to simulate a conventional data federation.  

\subsection{Workloads}
\textbf{HealthLNK:}
 We test \sysname over electronic health records from the \textit{HealthLNK} data repository~\cite{healthlnkirb}.  This clinical data research contains records from seven healthcare sites. The data repository contains about six million electronic health records from a diverse institutions\textendash including research hospitals, clinics, and a county hospital\textendash from 2006 to 2012. This dataset has
 significant skew, for example, a patient $A$ with a disease $X$, might have more vital recordings than patient $B$ with disease $Y$. Running \sysname on this datatset enables us to stress our model in the presence of significant skew. We map each site in the federation to a machine in our four-node testbed.   

We experiment with queries that are based on real clinical data research protocols for c.\@ diff infections and heart disease~\cite{hernandez2015adaptable,cdiff2015irb}. We use public patient registries for common ailments to bound the duration of our experiments.  A registry lists the patient identifiers associated with a condition with no additional information about the individual.   We maintain a patient registry for heart disease sufferers (\textit{hd\_cohort}) and one for individuals affected by  c.\@diff (\textit{cdiff\_cohort}), an infection that is frequently antibiotic-resistant.  Our queries are shown in Table~\ref{tbl:query-workload}.


\shortsection{TPC-H:} TPC-H is a standard synthetic workload and dataset which simulates an OLAP environment ~\cite{tpch}.  
We choose different scale factors depending on the specific experimental setup, using varying scale factors depending on the experiment.  We run use queries 3,5, and 10.

\begin{table*}
\scriptsize
\centering
\begin{tabular}{|l|l|}
\hline 
{\bf Name}  & {\bf Query} \\
\hline
{\it aspirin} & {\tt\scriptsize SELECT gender,  race, avg(pulse) FROM demographics de, diagnosis di, vitals v, medications m}\\
{\it profile} &   {\tt\scriptsize   WHERE  m.med = 'aspirin'  $\land$ di.diag = 'hd'  $\land$ dd.pid = di.pid $\land$ di.pid = v.pid $\land$ m.pid = di.pid; } \\
\hline
{\it comorbidity} & {\tt\scriptsize SELECT diag, COUNT(*) cnt FROM diagnoses WHERE pid $\in$ cdiff\_cohort $\land$ diag $<>$ 'cdiff' ORDER BY cnt  DESC LIMIT 10;} \\
\hline

	{\it dosage study} & {\tt\scriptsize SELECT pid FROM diagnoses d, medications m WHERE d.pid = m.pid AND medication = 'aspirin' AND}\\
	& {\tt\scriptsize icd9 = 'internal bleeding' AND dosage = '325mg'}\\
	\hline
\end{tabular}
\caption{HealthLNK query workload.} 
\label{tbl:query-workload}
\vspace{-3mm}
\end{table*}



\subsection{TPC-H}

\shortsection{Anonymized View Generation Scalability}

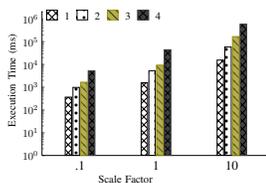
\begin{figure}[h]
    \centering
\resizebox{0.2\textwidth}{!}{

\begin{tikzpicture}
    \begin{axis}[
		    yscale=.45,
		    xscale=.6,
        width = 40em,
        major x tick style = transparent,
	ybar,
        bar width=1em,
        ylabel style={align=center,at={(axis description cs:-.05,0.85)}},
    ylabel={\,\, \,\,\, \,\,\, \,\, \,\,\,\,Execution Time (ms)},
    symbolic x coords={.1,1,10},        
        xtick = data,
        area legend,  
        enlarge x limits=0.25,
        ymin=1,
        ymode=log,
        axis y line*=left,
        axis x line*=bottom,
        x tick label style={font=\large,text width=2cm,align=center},
        xlabel={Scale Factor},
        legend cell align=left,
        legend style={
                at={(.8,2.0)},
                anchor=south east,
                column sep=1ex,
                legend columns=4,
                draw=none
        }
    ]

	    \addplot [pattern=crosshatch] coordinates {(.1,365) (1,1583) (10,15545)}; 
	    \addplot [pattern=dots] coordinates {(.1, 978.9) (1, 5208) (10, 58253)}; 
	\addplot [draw=olive, preaction={fill=olive}, pattern=north west lines, opacity=.7] 
	    coordinates {(.1, 1643) (1, 9250) (10, 165994)}; 
	\addplot [draw=darkgray, preaction={fill=darkgray},pattern=crosshatch] 
	    coordinates {(.1, 5114) (1, 43908) (10, 581558)}; 
    \legend{1,2,3,4}
\end{axis}
\end{tikzpicture}
}
    \vspace{-3mm}
	\caption{TPC-H Anonymization(SGX)}
	\label{fig:anon-scale-tpch}
     \vspace{-3mm}
\end{figure}

We run anonymized view generation on one, two, three, and four relations in SGX with four data owners. We scale the data size with TPC-H scale factors .1, 1, and 10. 
We use the customers, orders, lineitem, and supplier relations from the TPC-H schema. 
Orders is anonymized  on the "(o\_custkey)", lineitem on "(l\_suppkey,\\l\_orderkey)", supplier on "(s\_suppkey)", and customer on "(c\_custkey)"
The results are presented in Figure ~\ref{fig:anon-scale-tpch}.

The anonymization time scales roughly linearly with the data size: with a scale factor of .1 is 5s, with scale factor 1 is 44s , with scale factor 10 is 580s. 
The anonymization time is not uniform across relations, depending on data size and range.
 Gathering histograms requires running a $COUNT (*)$ type query on the relation, where the runtime will depend on size and range.  For example, processing the lineitem relation takes 60\% - 70\% of the time of the overall anonymization. 
Anonymization has substantial network costs since both the histograms have to be gathered at the coordinator, and then the anonymization 
maps must be distributed to all hosts. 

The greedy algorithm we implement for generating the $k$-anonymous processing views takes approximately 25\% of the time of 
anonymization. In workload mode anonymized views are retained and cached between queries, and thus the anonymization time can 
be amortized over a query workload. While the anonymization time for scale factor 10 is quite high, it is proportionally to query runtime on a dataset that large. 
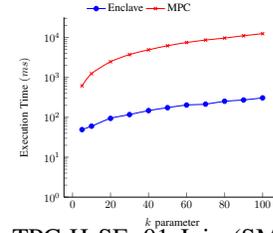
\begin{figure}[h]
    \centering
\resizebox{0.2\textwidth}{!}{
        \begin{tikzpicture}
 \begin{axis}[
        xlabel=$k$ parameter,
	 ylabel=Execution Time $(ms)$,
	ymode=log,
	ymin=1,
    ytick pos=left,
    xtick pos=left,
    axis y line*=left,
    axis x line*=bottom,
	 legend style={at={(.35, 1.1)}, anchor=north,legend columns=2, draw=none}
	 ]
    \addplot[smooth,mark=*,blue] plot coordinates {
	(5,49.233025)
	(10,59.87568)
	(20,93.90222)
	(30,117.0086)
	(40,147.28)
	(50,174.3158)
	(60,202.7354)
	(70,213.5396)
	(80,252.1468)
	(90,271.431)
	(100,306.6516)
    };
    \addlegendentry{Enclave}

    \addplot[smooth,color=red,mark=x]
        plot coordinates {
	(5,614.415083333)
	(10,1238.22)
	(20,2466.7475)
	(30,3723.40333333)
	(40,4901.10166667)
	(50,6204.863125)
	(60,7460.58)
	(70,8606.49833333)
	(80,9656.32166667)
	(90,11048.575)
	(100,12457.2833333)
        };
    \addlegendentry{MPC}
    \end{axis}
\end{tikzpicture}}
    \vspace{-3mm}
	\caption{TPC-H SF .01 Join (SMC, SGX)}
	\label{fig:query9_join}
     \vspace{-3mm}
\end{figure}

\shortsection{$K$-anonymous Join Study}
In this experiment we run a join on two hosts
with TPC-H tables $lineitem$ and $orders$, at scale factor .01. We anonymize the input relations with respect to the join key, \textit{orderkey}. The experiment includes both SMC as well as SGX execution.
As the $k$-anonymous parameter increases, join execution time increases proportional 
to the $k$ parameter. For a single join, we achieve nearly linear performance 
degradation as a parameter of $k$. We realize this very efficient result with the following intuition: given $n$ input tuples in each relation, and an anonymization parameter of $k$, each matched \ec produces O($k^2$) tuples.  The anonymized view generator produces approximately $n/k$ {\ec}es per relation. Hence, the output size of a $k$-anonymous join is $O(nk)$. The join's execution time is proportional to the size of its inputs, therefore as $k$ increases, we see a commensurate linear increase in execution time.

\textit{SMC: } In SMC, the data and anonymization are secretly shared across the two hosts. KPQ decreases amount of SMC comparisons
executed for the join predicate, since tuples are only compared within equivalence classes. As we increase $k$, those comparisons increase linearly, leading to the linear performance curve. With $k=5$, the SMC join takes 614ms, $k=100,$ 12457ms - a 20X performance penalty.

\textit{SGX:} The SGX backend, similar to SMC, has performance linear with $k$. However, SGX does not face the same steep performance 
degradation as SMC. With $k=5$, the SGX join takes 49ms, $k=100,$ 306ms - a 6.2X performance penalty. The SGX performance curve is linear 
with respect to $k$. 
\begin{figure}[h]
    \centering
\resizebox{0.3\textwidth}{!}{

\begin{tikzpicture}
    \begin{axis}[
        yscale=0.6,
        major x tick style = transparent,
	ybar,
        bar width=.5em,
        ylabel style={align=center},
    ylabel={\,\,\, \,\,\, \,\, \,\,\,\,Execution Time (ms)},
    symbolic x coords={Q3,Q5,Q10},        
        xtick = data,
        area legend,  
        enlarge x limits=0.25,
        ymin=1,
        ymode=log,
        axis y line*=left,
        axis x line*=bottom,
        x tick label style={font=\large,text width=2cm,align=center},
        xlabel={Query},
        legend cell align=left,
        legend style={
                at={(1,1.75)},
                anchor=south east,
                column sep=1ex,
                legend columns=4,
                draw=none
        }
    ]
	\addplot [pattern=crosshatch] coordinates {(Q3, 414) (Q5,857)(Q10,604)}; 
	\addplot [pattern=dots] coordinates {(Q3, 1402) (Q5, 1686) (Q10, 2635)}; 
	\addplot [draw=olive, preaction={fill=olive}, pattern=north west lines, opacity=.7] coordinates {(Q3, 1874) (Q5, 1817) (Q10, 3296)}; 
	\addplot [draw=blue, preaction={fill=blue}, pattern=north west lines, opacity=.7] coordinates {(Q3, 2093) (Q5, 2037) (Q10, 3950)}; 
	\addplot [draw=green, preaction={fill=green}, pattern=north west lines, opacity=.7] coordinates {(Q3, 2463) (Q5, 2124) (Q10, 5015)}; 
	\addplot [draw=green, preaction={fill=green}, pattern=north west lines, opacity=.7] coordinates {(Q3, 3302400) (Q5, 938814) (Q10, 3021245)}; 
    \legend{Encrypted, $5$-anon,$10$-anon,$15$-anon, $20$-anon, Oblivious}
\end{axis}
\end{tikzpicture}

}
    \vspace{-3mm}
	\caption{TPC-H Workload, SF .01(SGX)}
	\label{fig:tpch}
     \vspace{-3mm}
\end{figure}
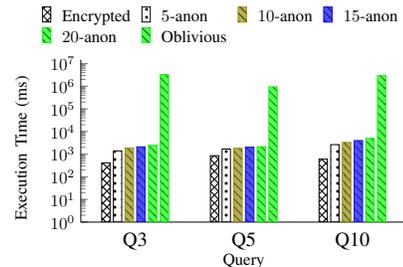

\shortsection{TPC-H Query Workload}
We run KloakDB on three TPC-H workload queries 3,5,10 in SGX with four data owners.
We run with scale factor $.01$ in order to allow the oblivious baseline to complete.
The experiment is run in encrypted mode, $k$-anonymous mode with $k=5,10,15, 20$, and oblivious mode. 
This experiment demonstrates two properties of \sysname: 1) Tunable performance with $k$, 2) performance over oblivious baseline. 
In 20-anon mode, the queries achieve the following performance improvements over 
the oblivious query processing: Q3,1340X; Q5,442X; Q10,602X. On the other hand, the performance penalty 
for 20-anon mode vs encrypted are the following: Q3,6X; Q5, 2.5X; Q10, 9X. 
Additionally, as we increase the $k$ from 5-20, the performance scales roughly linearly 
with the security parameter.

\subsection{HealthLNK Query Workload}
We run KloakDB on our real-world workload in SGX with four data owners, omitting the anonymization setup time in the presentation.  For the four relations used in our queries, the anonymization 
time for a year of data took a little over two seconds. Our experiments validate the viability of KQP on a dataset with significant skew.

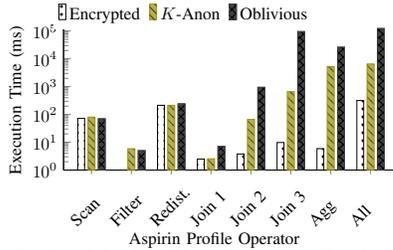
\begin{figure}[h]
    \centering
\resizebox{0.3\textwidth}{!}{
        \begin{tikzpicture}[
		every axis/.style={
		yscale=.56,
    ybar,
    axis y line*=left,
    axis x line*=bottom,
	legend style={at={(.755,1.95)},legend columns=3, draw=none},
    xlabel={},
    area legend,
	ylabel={Execution Time (ms)},
		    ylabel style={at={(axis description cs:0,.8)},anchor=north},
    xlabel style={at={(axis description cs:0.5,.05)},anchor=north},
    symbolic x coords={Scan, Filter, Redist., Join 1,Join 2,Join 3, Agg, All},
    ymode=log,
    log basis y=10,
    ymin = 1,
    xtick=data,
    bar width=.40em,
    xticklabel style={text width=3em, rotate=45},
    ytick pos=left,
		xtick pos=left
	},
    ]

	\begin{axis}[
			ylabel={},
			ticks=none,	
			symbolic x coords={},
		    xlabel style={at={(axis description cs:0.45,-.35)},anchor=north},
			xlabel={Aspirin Profile Operator},]
	\end{axis}
	
	\begin{axis}
	\addplot [pattern=dots] coordinates {(Scan, 72)  (Filter, 1) (Redist., 214) (Join 1, 2.5) (Join 2, 3.8) (Join 3, 9.80) (Agg, 5.9) (All, 318.2)};
	\addplot [draw=olive, preaction={fill=olive},pattern=north west lines, opacity=.7] coordinates {(Scan, 80) (Filter, 5.81) (Redist., 213) (Join 1, 2.53) (Join 2, 66.5) (Join 3, 660) (Agg, 5289) (All, 6516)};
	\addplot [draw=darkgray, preaction={fill=darkgray},pattern=crosshatch] coordinates {(Scan, 70) (Filter, 5) (Redist., 240.9) (Join 1, 7.07) (Join 2, 948) (Join 3, 95645) (Agg, 26238) (All, 123399)};
\legend{Encrypted\,\,,$K$-Anon\,\,, Oblivious}
\end{axis}
\end{tikzpicture}}
    \vspace{-3mm}
	\caption{Aspirin Profile, Operator Perf. (SGX)}
	\label{fig:query4}
     \vspace{-3mm}
\end{figure}

\shortsection{Aspirin Profile Operator Performance}
\label{sec:asp-profile}
 We analyze the per operator overhead of \sysname with the \textit{aspirin profile} query in Figure~\ref{fig:query4}.  We measure this query's runtime in encrypted, $k$-anonymous ($k=5$), and oblivious  mode. We randomly select 25 patients from the \textit{HealthLNK} dataset from one year of data.  

Figure~\ref{fig:query4} presents the operator runtime in each execution mode.   
The sequence of three joins is where KQP assumes a substantial performance gains over oblivious query processing.  In oblivious mode first join emits $n^2$ tuples, the second produces $n^3$, and so on.  In contrast, the expected cardinality of the first KQP join output is $O(nk)$ tuples, the second join $O(nk^2$) and so on.  The third join takes approximately 6 ms in encrypted mode, 650 ms in $k$-anonymous, and 93000 ms for oblivious processing.  This is a 103x slowdown between $k$-anonymous and encrypted, and a 143x slowdown between oblivious and $k$-anonymous execution. The aggregate in encrypted mode takes approximately 5ms, in $k$-anonymous 6700ms, oblivious 27900ms. The performance gap between $k$-anonymous mode and encrypted mode is due an unoptimized implementation, however the $k$-anonymous aggregate is 5x faster than oblivious execution.

The overall runtime for encrypted execution, $k$-anonymous, and oblivious is 320 ms, 6600ms, 123,000 ms respectively.  The slowdown incurred by $k$-anonymous execution compared to encrypted execution is 21X, and the speedup of $k$-anonymous execution in comparison to oblivious execution is 18X. Due to the prohibitively expensive overhead of oblivious execution, we sampled only 25  patients for {\em aspirin profile}. As the data size increases, we expect the gap between oblivious and $k$-anonymous execution to widen.

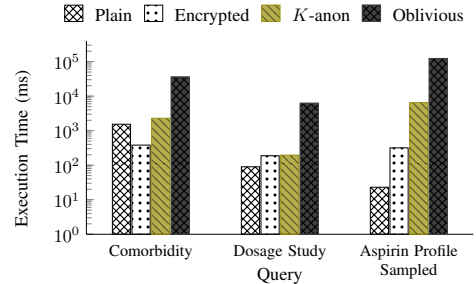
\begin{figure}[h]
    \centering
\resizebox{0.35\textwidth}{!}{
        \begin{tikzpicture}
    \begin{axis}[
        width  = 0.5*\textwidth,
        yscale=0.6,
        major x tick style = transparent,
        ybar=2*\pgflinewidth,
        bar width=1em,
        ylabel style={align=center},
    ylabel={\,\,\, \,\,\, \,\, \,\,\,\,Execution Time (ms)},
    symbolic x coords={Comorbidity,Dosage Study, Aspirin Profile Sampled},        
        xtick = data,
        area legend,  
        enlarge x limits=0.25,
        ymin=1,
        ymode=log,
        axis y line*=left,
        axis x line*=bottom,
        x tick label style={font=\small,text width=2cm,align=center},
        xlabel={Query},
        legend cell align=left,
        legend style={
                at={(1,1.75)},
                anchor=south east,
                column sep=1ex,
                legend columns=4,
                draw=none
        }
    ]
	\addplot [pattern=crosshatch] coordinates {(Comorbidity,1536) (Dosage Study,90)(Aspirin Profile Sampled,23)}; 
	\addplot [pattern=dots] coordinates {(Comorbidity, 382) (Dosage Study, 188) (Aspirin Profile Sampled, 318.2)}; 
	\addplot [draw=olive, preaction={fill=olive}, pattern=north west lines, opacity=.7] coordinates {(Comorbidity, 2281) (Dosage Study, 195) (Aspirin Profile Sampled, 6516)}; 
	\addplot [draw=darkgray, preaction={fill=darkgray},pattern=crosshatch] coordinates {(Comorbidity, 35830) (Dosage Study, 6235) (Aspirin Profile Sampled, 123399)}; 
    \legend{Plain,Encrypted, $K$-anon, Oblivious}
\end{axis}
\end{tikzpicture}

}
    \vspace{-3mm}
	\caption{HealthLNK Query Workload (SGX)}
	\label{fig:full_workload}
     \vspace{-3mm}

\end{figure}

\shortsection{Full Distributed Workload}
\label{sec:e2e-results}
In this section we run the full query workload in Table~\ref{tbl:query-workload}.  We demonstrate that $k$-anonymous query processing provides substantial performance improvements over oblivious query processing while providing data protection in comparison to encrypted execution. We run the queries in four modes: \textit{plain}, encrypted , $k$-anonymous, and oblivious. For the \textit{comorbidity} and \textit{dosage study} queries we run the queries on a full year of data in all four modes. However, the \textit{aspirin profile} queries was unable to complete in oblivious mode on a full year of data, therefore we sample 25 unique \textit{patients} per host.  
Figure ~\ref{fig:full_workload}  has the results of the full  workload.  The \textit{comorbidity} query demonstrates that even with a simple query, \kqp is an attractive alternative to oblivious query processing. $K$-anonymous execution has a 15X speedup compared to oblivious
execution, and 6X slowdown compared to encrypted.   
The \textit{dosage study} query sees a 31X speedup in $k$-anonymous execution compared to oblivious execution, and a 1.03X slowdown in $k$-anonymous execution compared to encrypted. We detailed the performance for \textit{aspirin profile} in this query in Section~\ref{sec:asp-profile}.

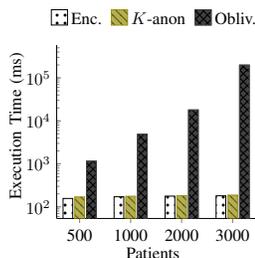
\begin{figure}[h]
    \centering
\resizebox{0.2\textwidth}{!}{
        \begin{tikzpicture}
\begin{axis}[
    yscale=.63,
    xscale=.6,
        ymode=log,
	ymin=0,
    ybar,
        area legend,
    enlargelimits=0.15,
    legend style={at={(0.75,2.2)},
      anchor=north,legend columns=3, draw=none},
      xlabel={Patients},
    xlabel style={at={(axis description cs:0.73,.28)},anchor=north},
ylabel={Execution Time (ms)},
    y label style={at={(axis description cs:.2,1.1)},anchor=south},
    symbolic x coords={500,1000,2000, 3000},
    xtick=data,
    bar width=.95em,
    ytick pos=left,
    xtick pos=left,
    axis y line*=left,
    axis x line*=bottom
    ]
	\addplot [pattern=dots] coordinates {(500,157) (1000,172) (2000, 178) (3000,181)};  
	\addplot [draw=olive, preaction={fill=olive}, pattern=north west lines, opacity=.7] coordinates { (500,170) (1000,176) (2000, 183) (3000, 187)};
	\addplot [draw=darkgray, preaction={fill=darkgray},pattern=crosshatch] coordinates { (500,1154) (1000,4876) (2000, 17988) (3000, 198369)};
    \legend{Enc.\,\,,$K$-anon\,\,, Obliv.}
\end{axis}
\end{tikzpicture}}
    \vspace{-3mm}
	\caption{Dosage Study Scale (SGX)}
	\label{fig:dosage_scale}
     \vspace{-3mm}

\end{figure}

\shortsection{Dosage Study Scale-up}
In this section we verify that as the input tuple size increases, the gap between $k$-anonymous execution 
and oblivious execution widens. We use  \textit{dosage study}  so that oblivious execution may complete.  We vary the number of patients we sample to  measure performance changes for data of increasing size in encrypted, $5$-anonymous, and oblivious mode.  
Figure ~\ref{fig:dosage_scale} shows the runtime of this query. $K$-anonymous execution is slightly slower than oblivious execution with 500 sampled patients owing to the overhead of its\kpv setup.  As the input size increases, $k$-anonymous query processing offers significant performance benefits over full-obliv\-ious query processing. With 3,000 patients, the runtime for encrypted, $k$-anon\-ymous, and full-ob\-livious query processing respectively are approximately 181ms, 187ms, and 198369. This yields 1.03X slowdown for $k$-anonymous mode compared to encrypted, and 1060X speedup for it in comparison to oblivious. The stark slowdown for oblivious mode is due to the substantial memory pressure imposed by exploding cardinalities, leading the join output to spill to disk one {\ec} at a time. In $k$-anonymous mode, scaling the input size from 500 to 3000 patients yields a 1.1X slowdown. This stands in contrast with the 171X slowdown we observe in oblivious mode. This experiment highlights an important feature of this system: \sysname enables substantial speedups for query processing as input data scales.

\section{Related Work}
\label{related}

{\sysname} builds on principles in secure query processing, oblivious computation, and secure computation. We survey the existing research in these areas. 

Private data federations and were

K-anonymity has been studied extensively. 
{these people} discuss how to maximize data utility with respect to 
some objective function. {mutiR paper} discusses an k-anonymization scheme 
for multiple relations. 
Most of this work discusses k-anonymity in the context of data release. T
Secure query processing has. K-anonymous query processing is heavily influenced 
by these previous works. 

{\sysname} builds on principles in query processing, applied security, and automated access control policies.   There is substantial active research in all of these areas and we survey them in this section.

Speaking broadly, there are two common methods for methods for general-purpose computing over the data of two or more mutually distrustful parties: in software with secure multi-party computation~\cite{Goldreich1987,yao1982protocols} and in hardware using hardware enclaves~\cite{costan2016intel,kaplan2016amd}.  The former is possible on any system, but exacts a substantial overhead in making the computation oblivious and encrypting its contents.   The latter requires specialized hardware, but is more efficient.   We chose hardware enclaves for this work, and the principles of \kqp  readily generalize to secure multi-party computation.

There has been substantial work on oblivious query processing using hardware enclaves~\cite{arasu2014oblivious, bajaj2011trusteddb, eskandarian2017oblivious, priebe2018enclavedb, zheng2017opaque, Ohrimenko2016OMPCML}.  In this setting a curious observer of an enclave  learns nothing about the data upon which they compute by observing its instruction traces.  We build on this work by offering semi-oblivious query processing for querying data of moderate sensitivity.

\sysname is a \federation.  This challenge was researched with the use of secure multi-party computation to combine the private data of multiple parties in~\cite{bater2017smcql,bogdanov2008sharemind,conclave,wong2014secure}.  We extend this work, but examine how to do it semi-obliviously\textendash rather than with full guarantees of cryptographic hardness\textendash  in exchange for faster query runtimes.  
Shrinkwrap ~\cite{bater2019shrinkwrap} considers a similar semi-oblivious model through reducing the output cardniality 
of joins with differential privacy. Our work differs in two ways: 1) Shrinkwrap still 
requires executing the full cross product for joins and only after runs the Shrinkwrap protocol, 2)
Shrinkwrap does not support multiple queries. 

$K$-anonymous data releases were proposed in~\cite{samarati1998protecting}.  There has been substantial work on efficiently generating $k$-anony\-mous views of a given dataset~\cite{
bayardo2005data,el2009globally,Jiang2006,lefevre2005incognito,lefevre2006mondrian,sweeney2002k}.  \sysname extends the techniques in~\cite{doka2015k} to build {\kpv}s.  We generalize the requirements of multi-relational $k$-anonymous data releases in ~\cite{nergiz2009multirelational} to \sysname's computational model.  
Automatically enforcing $k$-anon\-ymous access control policies in a dataset was researched in~\cite{eltabakh2012query}.  

Most of the prior work on oblivious query processing focuses on outsourced computation from a single data provider, either in software with secure multi-party computation~\cite{aggarwal2005two} or in hardware with hardware enclaves~\cite{ arasu2014oblivious, zheng2017opaque}.  Some of them~\cite{bater2017smcql,bogdanov2008sharemind,conclave, wong2014secure} offer interoperability for multiple {\owner}s. 

There has been limited work on semi-oblivious computation.  The most common method for this is computational differential privacy~\cite{mironov2009computational}.  Protocols of this kind leak noisy information about the data and they are analogous to computing on a differentially private version of the dataset.



There is also work about computing queries in the cloud over data stored with fully homomorphic encryption~\cite{Popa2011, tu2013processing}.  Encrypted databases have reduced expressiveness since they cannot readily compose operators for nested blocks of select statements.  Because \sysname protects the query's computation instead of the data, it supports nested queries.



\section{Conclusions}
\label{conclusion}
We presented a semi-oblivious query processing model, \kqp for private data federations. With KQP, data owners have a fine tuned knob with which 
to trade off privacy and performance. Our formalization of \kqp allows for complex queries, while protecting against unauthorized privacy leakage 
in the multiple query setting. This is an important step towards more approachs that strike a balance between security and performance for querying private data. Our model is grounded in $k$-anonymity, a relevant and widely deployed privacy model.  We built and tested a prototype \sysname, utilizing both secure multiparty computation and hardware enclaves to provide confidentiality. 

Our evalution shows KQP provides fine-grained tunablility for increasing privacy. Our join study demonstrates a linear tradeoff between privacy and performance. We demonstrate on TPC-H workloads speedups of up to 440X-2350X. On a real-world dataset with a real-world workload we demonstrate speedups of 15X-1060X.
Our results show that if KQP is the appropriate query processing model for a data federation, there is significant room for performance and scalability gains.



\bibliographystyle{IEEEtran}
\bibliography{main}

\end{document}